\documentclass[prb,11pt]{revtex4-1}

\usepackage{amsmath}    
\usepackage{graphicx}   
\usepackage{verbatim}   
\usepackage{color}      
\usepackage{subfigure}  

\begin{document}

\title{
 Improving the Description of Nonmagnetic and Magnetic Molecular Crystals via the van der Waals Density Functional
}


\author{Masao Obata}
\affiliation{Graduate School of Natural Science and Technology, Kanazawa University, Kakuma, Kanazawa 920-1192, Japan}

\author{Makoto Nakamura}
\affiliation{Graduate School of Natural Science and Technology, Kanazawa University, Kakuma, Kanazawa 920-1192, Japan}

\author{Ikutaro Hamada}
\affiliation{International Center for Materials Nanoarchitectonics (WPI-MANA), National Institute for Materials Science (NIMS), Tsukuba 305-0044, Japan}

\author{Tatsuki Oda}
\affiliation{Graduate School of Natural Science and Technology, Kanazawa University, Kakuma, Kanazawa 920-1192, Japan}
\affiliation{Institute of Science and Engineering, Kanazawa University, Kakuma, Kanazawa 920-1192, Japan}

\begin{abstract}
  We have derived and implemented a stress tensor formulation for the van der Waals density functional (vdW-DF) with spin-polarization-dependent gradient correction (GC) recently proposed by the authors [J. Phys. Soc. Jpn. {\bf 82}, 093701 (2013)] and applied it to nonmagnetic and magnetic molecular crystals under ambient condition.
  We found that the cell parameters of the molecular crystals obtained with vdW-DF show an overall improvement compared with those obtained using local density and generalized gradient approximations.
  In particular, the original vdW-DF with GC gives the equilibrium structural parameters of solid oxygen in the $\alpha$-phase, which are in good agreement with the experiment.
\end{abstract}

\maketitle

%
\section{Introduction \label{introduction}}
%
%
The oxygen molecule has a spin-triplet ground state, and the competing magnetic and molecular interactions
result in a rich variety of structural, electronic, and magnetic phases under wide temperature and pressure ranges.\cite{Freiman2004}
Much effort has been devoted to clarifying the magnetism and atomic structure of solid oxygen.\cite{DeFotis1981, Klotz2010, Meier1984, Etters1985, Otani1998, Kususe1999, Nozawa2002, Nozawa2008}
The magnetism of solid oxygen is particularly interesting,  as the magnetic moment is sensitive to the relative distance between molecules owing the short-range feature of the direct magnetic interaction.\cite{Santoro2001}
Thus, accurate prediction of the crystal structure is one of the key issues for understanding the magnetism in  solid oxygen.
Theoretical works on solid oxygen in the literature are mostly for those under a high pressure, presumably because under a relatively low pressure, the vdW interaction plays an important role, where the conventional semilocal density functional theory (DFT) fails to describe it properly.
An accurate description of the vdW interaction is, however, challenging when using the electronic structure method, particularly, DFT within the semilocal approximation, which is known to describe the dispersion interaction poorly.
In quantum chemistry, there are well-established methods, such as the second-order M{\o}ller-Plesset perturbation theory (MP2), coupled-cluster calculations with single, double, and perturbative triple excitations [CCSD(T)], and the symmetry-adopted perturbation theory (SAPT), which can describe the vdW interaction.
The quantum Monte-Carlo (QMC) method and the adiabatic connection fluctuation dissipation theorem within the random phase approximation (ACFDT-RPA) are also known to describe the vdW interaction accurately.
However, because these highly accurate methods are computationally very demanding and applicable only to small systems, practical methods based on DFT are desirable for the simulations of complex systems.
In this sense, semi-empirical dispersion-corrected DFT\cite{Grimme2006} is widely used, and recent developments\cite{Grimme2010,Tkatchenko2009,Tkatchenko2012} allow more accurate calculations of the vdW interaction with an almost negligible overhead.
A more elaborate approach based on the maximally localized Wannier function was also proposed to correct for the missing dispersion interaction in DFT.\cite{Silvestrelli2008,Silvestrelli2013}
Among other approaches, the van der Waals density functional (vdW-DF), which was developed by Rydberg {\it et al.}\cite{Rydberg2003} and Dion {\it al.},\cite{Dion2004} has attracted much attention, because it does not require empirical parameters and depends only on charge density and its gradient, thereby allowing one to perform large-scale calculation at a modest computational cost.
There have been applications of vdW-DF to several systems, and to extend the applicability of the methods, there have been several proposals for more accurate vdW-DFs.\cite{Vydrov2009, Klimes2010,Cooper2010, Lee2010, Vydrov2010, Klimes2011,Wellendorff2012,Sabatini2013,Berland2014, Hamada2014}
However, because the vdW interaction is not directly related to spin polarization, a truly spin-polarized version of vdW-DF has not yet been developed, except for the one proposed by Vydrov and Van Voorhis.\cite{Vydrov2009}
Thus, the application of vdW-DFs to magnetic systems is quite limited.
In our previous work,\cite{Obata2013} we proposed a practical approach to a magnetic system within the framework of vdW-DF, in which spin-polarization-dependent gradient correction is added to the local correlation energy and potential, instead of developing the spin-polarized version of the nonlocal correlation.
The method has been applied to an oxygen dimer, and shown to be applicable to magnetic systems.
In this work, we have studied solid oxygen at ambient pressure in the $\alpha$ phase ($\alpha$-O$_{2}$) using our vdW-DF for magnetic materials.
We implemented the stress tensor arising from the nonlocal correlation to optimize the cell parameters of $\alpha$-O$_{2}$ and used several vdW-DFs with different exchange and nonlocal correlation.
We found that one of the vdW-DFs shows a significant improvement on the structural properties over the conventional local density approximation (LDA) or generalized gradient approximation (GGA).
The remaining part of this paper is organized as follows.
In Sect.~\ref{sec:method}, we describe our implementation of vdW-DF and the computational detail.
To test our implementation of vdW-DF, we first applied our code to solid argon, graphite, selenium, and dry ice (solid carbon dioxide), in which vdW forces play an important role in the binding, and the results are given in Sects.~\ref{secsec:ar}--\ref{secsec:co2}.
The results for solid oxygen are discussed in detail in Sect.~\ref{secsec:o2}.
The summary for this paper is given in Sect.~\ref{sec:summary}.
\section{Method \label{sec:method}}
%
\subsection{van der Waals density functional}
%
The exchange-correlation energy in vdW-DF is given\cite{Dion2004} by
\begin{align}
  E_{\rm xc} = E_{\rm x} + E_{\rm c}^{\rm loc} + E_{\rm c}^{\rm nl},
  \label{eq:exc}
\end{align}
where $E_{\rm x}$, $E_{\rm c}^{\rm loc}$, and $E_{\rm c}^{\rm nl}$ are the exchange energy, short-range local correlation energy, and nonlocal correlation energy, respectively.
$E_{\rm c}^{\rm nl}$ is expressed as a double integral of the nonlocal interaction kernel 
$\phi(q_1,q_2,r_{12})$\cite{Dion2004} over the spatial coordinates ${\bf r}_1$ and ${\bf r}_2$ as
\begin{align}
  E_{\rm c}^{\rm nl} =
   \frac{1}{2} \iint n({\bf r}_1) \phi(q_1, q_2, r_{12}) n({\bf r}_2) d{\bf r}_1d{\bf r}_2 ,
   \label{eq:ecnl-r-sum}
\end{align}
where $r_{12}=|{\bf r}_1 - {\bf r}_2|$, 
$q_1=q_0({\bf r}_1)$, $q_2=q_0({\bf r}_2)$, and $q_0({\bf r})$ is a function of the charge density $n({\bf r})$ and its gradient $|\nabla n({\bf r})|$. 
Rom\'{a}n-P\'{e}rez and Soler (RPS)\cite{Roman-Perez2009} proposed an efficient method for evaluating $E_{\rm c}^{\rm nl}$, as the direct computation of the double integral in Eq.~(\ref{eq:ecnl-r-sum}) is extremely time-consuming.
In the RPS scheme, the vdW kernel $\phi$ is expanded in terms of the interpolating function $p_{\alpha}(q)$, which satisfies $p_{\alpha}(q_{\beta})=\delta_{\alpha \beta}$, so that $\phi$ is a fixed value at a given $q_{\beta}$ point, allowing the use of the fast Fourier transform (FFT) in the evaluation of $E_{\rm c}^{\rm nl}$.
However, $\phi$ has a divergence when $q_{\alpha}, q_{\beta} \rightarrow 0$, which makes the interpolation near the origin difficult, and thus RPS replaced $\phi$ with the soft form, and an LDA-like approximation was employed for $E_{\rm c}^{\rm nl}$ near the origin.
Alternatively, Wu and Gygi (WG)\cite{Wu2012} proposed a simplified implementation to avoid the divergence in $\phi$ as follows.
The nonlocal interaction kernel, multiplied by $q_1$ and $q_2$, is expanded as
\begin{align}
  q_1 q_2 \phi(q_1,q_2,r_{12}) \simeq \sum_{\alpha\beta}  q_\alpha p_{\alpha}(q_1) q_\beta p_\beta(q_2)\phi_{\alpha \beta}(r_{12}),
  \label{eq:phi}
\end{align}
so that the divergence at $q_{\alpha}, q_{\beta} \rightarrow 0$ can be avoided.
By introducing the function
$\eta_{\alpha}({\bf r}) = q_{\alpha} n({\bf r }) p_{\alpha} (q_0({\bf r}))/q_0({\bf r)}$,
the nonlocal correlation is calculated in the reciprocal space as
\begin{align}
  E_{\rm c}^{\rm nl} &= \frac{1}{2} \sum_{\alpha \beta} 
                         \iint  \eta_\alpha({\bf r}_1) 
                         \eta_\beta ({\bf r}_2) \phi_{\alpha \beta}(r_{12}) d{\bf r}_1 d {\bf r}_2 \nonumber \\
                     &= \frac{\Omega}{2}\sum_{\bf G} \sum_{\alpha \beta} 
                         \eta^*_\alpha({\bf G}) \eta_{\beta}({\bf G}) 
                         \phi_{\alpha \beta}(G), 
  \label{eq:ecnl-g-sum}
\end{align}
where ${\bf G}$ is the reciprocal lattice vector $G=\vert {\bf G} \vert$, $\eta_{\alpha}({\bf G})$ and $\phi_{\alpha\beta}(G)$ are the Fourier components of 
$\eta_{\alpha}({\bf r})$ and $\phi_{\alpha\beta}(r_{12})$, respectively, and $\Omega$ is the volume of a unit cell. 
The Fourier component $\phi_{\alpha \beta}(G)$ is calculated as
\begin{align}
  \phi_{\alpha \beta}(G) = \frac{1}{q_{\alpha}^3} F_{\alpha \beta} \left( \frac{G}{q_{\alpha}} \right),
\label{eq:phi-alpha-beta-g}
\end{align}
where
\begin{equation}
F_{\alpha \beta} (k) = \frac{4 \pi}{k} \int_0^{\infty} u \phi \left(u, \frac{q_{\beta}}{q_{\alpha}}u \right) \sin(k u) du .
\label{eq:f-alpha-beta-k}
\end{equation}
Note that the ${\bf G}=0$ term in Eq.~(\ref{eq:ecnl-g-sum}) vanishes when $\alpha=\beta$. 
We discuss the behaviors of the $\phi$ function in Appendix~\ref{appendix1}.

The nonlocal correlation potential within the White-Bird algorithm\cite{White1994} is given by
\begin{align}
v_{\rm c}^{\rm nl}({\bf r}_i)
         &=
         \frac{ \delta E_{\rm c}^{\rm nl}}{ \delta n({\bf r}_i)}  \nonumber \\
         &\simeq
         \frac{N}{\Omega} \frac{d E_{\rm c}^{\rm nl}}{d n({\bf r}_i)} \nonumber \\
         &=
         \sum_\alpha \left(  u_{\alpha }({\bf r}_i) 
         \frac{\partial \eta_{\alpha }({\bf r}_i)}{\partial n({\bf r}_i)} 
         + \sum_j u_{\alpha }({\bf r}_j)
           \frac{\partial \eta_{\alpha }({\bf r}_j)}{\partial \nabla n({\bf r}_j)}
           \frac{\partial \nabla n ({\bf r}_j)}{\partial n ({\bf r}_i)} \right),
\label{eq:v-ecnl}
\end{align}
where 
\begin{align}
  u_{\alpha }({\bf r}_i) = \frac{\Omega}{N} \sum_{\beta j} 
  \phi_{\alpha \beta}(r_{ij}) \eta_{\beta }({\bf r}_j)= \sum_{ \beta \bf G} \phi_{\alpha \beta}(G) \eta_{\beta}({\bf G}) e^{i {\bf G} \cdot {\bf r}_i},
    \label{eq:ecnl-u}
\end{align}
with $N$ being the number of FFT grids.
%

\subsection{Pressure tensor}

To optimize the lattice parameter, it is necessary to calculate the pressure tensor, and there is an additional contribution from the nonlocal correlation in vdW-DF.\cite{Sabatini2012, Wu2012}
In our formulation, an additional contribution to the pressure tensor $\Pi_{k\ell}, (k,\ell=x,y,z)$ is calculated as
\begin{align}
  (\Pi_{\rm c}^{\rm nl})_{k\ell} =& 
    -\frac{1}{\Omega} \sum_{m} \frac{\partial E_{\rm c}^{\rm nl}}{\partial h_{km}} (h^t)_{m\ell} \\
    =& -\frac{1}{\Omega} E_{\rm c}^{\rm nl} \delta_{k\ell} \nonumber \\
     & - \frac{1}{N} \sum_{m} \sum_{j} \sum_{\alpha}  u_{\alpha}({\bf r}_j) 
    \frac{\partial \eta_{\alpha} ({\bf r}_j)}{\partial n({\bf r}_j)}
    \frac{\partial n({\bf r}_j)}{ \partial h_{km}} (h^t)_{m\ell}  \nonumber \\
& + \frac{1}{N} \sum_{j} \sum_{\alpha} u_{\alpha}({\bf r}_j)  
    \frac{\partial \eta_{\alpha} ({\bf r}_j)}{\partial |\nabla n({\bf r}_j)|} 
    \frac{(\nabla n({\bf r}_j) )_k (\nabla n({\bf r}_j))_{\ell}}
           {|\nabla n({\bf r}_j)|}  \nonumber \\
& - \frac{1}{N} \sum_{m} \sum_{j} \sum_\alpha u_{\alpha}({\bf r}_j) 
    \frac{\partial \eta_{\alpha} ({\bf r}_j)}{ \partial |\nabla n({\bf r}_j)|}   \nonumber \\
&\quad \times \frac{\nabla n({\bf r}_j)}{ |\nabla n({\bf r}_j)|}  \cdot \sum_{\bf G}
    \frac{\partial n({\bf G})}{ \partial h_{km} } (h^t)_{m\ell} 
    (i {\bf G}) e^{i {\bf G} \cdot {\bf r}_j} \nonumber \\
& + \frac12 \sum_{\alpha \beta} \sum_{\bf G} \eta_{\alpha}^*({\bf G}) \eta_{\beta}({\bf G}) 
    \frac{\partial \phi_{\alpha\beta} (G)}{\partial |{\bf G}|}
    \frac{G_k G_\ell}{|{\bf G}|}, 
\label{eq:p-ecnl}
\end{align}
where the matrix $h=({\bf a}_1,{\bf a}_2,{\bf a}_3)$ is a set of 
primitive lattice vectors ${\bf a}_k$ $(k=1,2,3)$ and $h^t$ is the inverse of $h$.
%
\subsection{Spin-polarization-dependent gradient correction of the local correlation}
%
In the original vdW-DF, LDA was employed for the short-range local correlation energy, to avoid possible double counting of the contribution  from $\vert \nabla n \vert$ contained in $E_{\rm c}^{\rm nl}$.\cite{Dion2004}
In a spin-polarized system, spin-polarization-dependent gradient correction should be included in the correlation energy and potential, but such a contribution is missing in the nonlocal correlation, as it is mostly formulated for the spin-unpolarized system.
In our previous work\cite{Obata2013}, we propose a simple scheme to include such spin-polarization-dependent gradient contribution as follows.
We define the local part of the correlation functional as
\begin{align}
E_{\rm c}^{\rm loc}= E_{\rm c}^{\rm LSDA}[n_{\uparrow},n_{\downarrow}] + 
                     \Delta E_{\rm c}[n,\zeta],
\label{eq6}
\end{align}
where
\begin{align}
\Delta E_{\rm c}[n,\zeta] &=
E_{\rm c}^{\rm GGA}[n_{\uparrow}, n_{\downarrow}] - 
E_{\rm c}^{\rm GGA}[n/2, n/2]  \nonumber \\
                          &=
\int n \left\{H(r_{s},\zeta ,t)-H(r_{s},0,t)\right\} d{\bf r}, 
\label{eq7}
\end{align}
$H$, $r_{\rm s}$, $\zeta$, and $t$ are the gradient contribution, Seitz radius ($n=3/4\pi r_{s}^{3}$), relative spin polarization, and dimensionless density gradient proportional to $|\nabla n|$, respectively.
We use the functional form for $H$ proposed by Perdew, Burke, and Ernzerhof (PBE).\cite{Perdew1996}
The present functional is reduced to the original one in the absence of spin polarization.
%

\subsection{Technicalities\label{Tech}}
%
We have implemented vdW-DF in our in-house DFT code,\cite{Car1985,Oda1998,Oda2002} which uses a plane-wave basis set and ultrasoft pseudopotentials.\cite{Vanderbilt1990,Laasonen1993}
In the construction of the pseudopotentials, we neglected the nonlocal correlation ($E_{\rm c}^{\rm nl}$) and employed the semilocal exchange and correlation functionals, which are consistent with the solid-state calculation.
For instance, the pseudopotential constructed using the refit Perdew Wang (PW86R)\cite{Perdew1986,Murray2009} exchange and the LDA correlation was used in vdW-DF2 calculation.
The use of the pseudopotential generated with the semilocal exchange correlation functional has been justified.\cite{Klimes2011, Hamada2014}
We also performed the calculations using the pseudopotential generated with the PBE functional and found that there is an insignificant difference in binding energy ($\sim$ 1~meV/atom).
In the present vdW-DF calculations, we used relatively large energy cutoffs for wave functions and electron density in the plane wave expansions, compared with those used in LDA or GGA, to achieve the convergence of the pressure tensor, because the vdW forces are very weak, and the potential energy surface is very sluggish.
Unless otherwise noted, we optimized the lattice parameters and associated internal structural parameters to converge within a pressure of 0.05 GPa.
The Brillouin zone integration was performed using the set of Monkhorst-Pack (MP) special $k$-points.\cite{Monkhorst1976} 
In this work, we employed the Perdew-Zunger exchange correlation for LDA \cite{Perdew1981} and the PBE functional for GGA\cite{Perdew1996}.
For the vdW-DF calculations, we employed the following functionals:
the original vdW-DF with the revPBE\cite{Zhang1998} exchange and LDA correlation, the second version of the vdW-DF (vdW-DF2), which uses PW86R\cite{Murray2009} exchange, vdW-DF paired with Cooper's (C09)\cite{Cooper2010} exchange (vdW-DF$^{\rm C09x}$), and vdW-DF2 paired with C09 exchange (vdW-DF2$^{\rm C09x}$).
We also implemented the revised Vydrov-Van Voorhis functional (rVV10)\cite{Vydrov2010,Sabatini2013} with the WG\cite{Wu2012} implementation (see Appendix \ref{appendix2}), which uses the PW86R exchange, PBE correlation, and the nonlocal correlation.
The functional is designed to give accurate $C_{6}$ coefficients and to vanish in the uniform electron gas limit.
In the case of antiferromagnetic $\alpha$-O$_{2}$, we employed the spin-polarization-dependent gradient correction (GC) as described previously, and ``-GC'' is appended to the functional name when GC is considered \cite{Obata2013}.
Note that GC is not necessary for rVV10, because the PBE-GGA correlation is used for the local correlation.
To evaluate $\phi$ in the vicinity of $q_{\alpha}=0$, we use a linear mesh up to $q=q_m$ with $N^{\rm lin}$ grid points, and the logarithmic mesh is used from $q_m$ to $q_c$ with $N^{\rm log}$ grid points, where $q_c$ is the cutoff for the $q$-mesh.
We use $q_c=8~{\rm a.u.}, q_m=0.01~{\rm a.u.}, N^{\rm lin}=3$, and $N^{\rm log}=28$, except for rVV10, in which  $q_c=3~{\rm a.u.}$, and $q_m=0.005~{\rm a.u.}$ are used.
%
We confirmed that the cutoff $q_{c}$ is sufficiently large, by monitoring the distribution of $q_{0}$ and the nonlocal correlation energy as a function of $q_{0}$ (see Appendix \ref{appendix3}).
%

%
\section{Results \label{sec:results}}
%
\subsection{Fcc and hcp argons \label{secsec:ar}}

\begin{table}[tb]
  \begin{center}
    \caption{Optimized lattice parameter ($a_{\rm fcc}$),
             equilibrium volume ($V_{0}$), and cohesive energy ($\Delta E_{\rm coh}$) for solid argon in the fcc structure, obtained using different density functionals. The CCSD(T), ACFDT-RPA using PBE orbitals, and experimental values are also listed for comparison.}
    \label{table:Ar-fcc}
    \begin{tabular}{lccc} \hline \hline
      Functional           & $a_{\rm fcc}$ & $V_0$             & $\Delta E_{\rm coh}$ \\
                           & (\AA)         & (\AA ${^3}$/atom) & (meV/atom)\\ \hline
      LDA                  & 4.97          & 30.61             & 135     \\
      PBE                  & 6.04          & 55.03             & 14     \\
      vdW-DF               & 5.53          & 42.34             & 142     \\
      vdW-DF2              & 5.29          & 37.05             & 116     \\
      vdW-DF$^{\rm C09x}$  & 5.34          & 38.06             & 105     \\
      vdW-DF2$^{\rm C09x}$ & 6.32          & 63.06             & 21     \\
      rVV10                & 5.20          & 35.20             & 99     \\
      rVV10$^{a}$          & 5.17          & 34.55             & 117 \\
      CCSD(T)$^{b}$        & 5.251         & 36.196            & 87.9 \\
      ACFDT-RPA$^{c}$      & 5.3           & 37.22             & 83 \\
      Expt.                & 5.311$^d$     & 37.451$^d$        & 80.1$^e$ \\ \hline \hline
    \end{tabular}
  \end{center}
  \begin{center}
    $^a$ Ref.~\onlinecite{Tran2013}. $^b$ Ref.~\onlinecite{Posciszewski2000}. $^c$ Ref.~\onlinecite{Harl2008}. $^d$ Ref.~\onlinecite{Peterson1966}.
    $^e$ Ref.~\onlinecite{Schwalbe1977}.
  \end{center}
\end{table}

\begin{table}[tb]
  \begin{center}
    \caption{Optimized lattice parameters ($a_{\rm hcp}$ and $c_{\rm hcp}$), equilibrium volume ($V_{0}$),  
             and cohesive energy ($\Delta E_{\rm con}$) for hcp argon.}
    \label{table:Ar-hcp}
    \begin{tabular}{lcccc} \hline \hline
      Functional    
      & $a_{\rm hcp}$      & $c_{\rm hcp}/a_{\rm hcp}$ & $V_0$             & $\Delta E_{\rm coh}$   \\
      & (\AA)              &                           & (\AA ${^3}$/atom) &  (meV/atom) \\ \hline
      LDA                  & 3.51        &  1.633   & 30.45    & 135  \\
      PBE                  & 4.16        &  1.633   & 50.92    &  13  \\
      vdW-DF               & 3.89        &  1.633   & 41.63    & 142  \\
      vdW-DF2              & 3.74        &  1.636   & 37.08    & 116  \\
      vdW-DF$^{\rm C09x}$  & 3.80        &  1.632   & 38.75    & 105  \\
      vdW-DF2$^{\rm C09x}$ & 4.19        &  1.632   & 52.81    &  20  \\
      rVV10                & 3.66        &  1.633   & 34.70    &  99  \\
      Expt.$^{a}$          & 3.8         &  1.63    & 39       &      \\ \hline \hline
    \end{tabular}
  \end{center}
  \begin{center}
   $^a$ Ref. \onlinecite{Wittlinger1997}.
  \end{center}
\end{table}

We begin with solid argon in the face-centered-cubic (fcc) and hexagonal-close-packed (hcp) structures, as a typical vdW bonded solid.
It is known that at ambient pressure, argon in the fcc structure is the most stable, whereas that in the hcp structure is metastable.
For both phases, we optimized lattice parameters using several vdW-DFs as well as LDA and GGA.
The energy cutoffs of 60 and 500 Ry were used for wave function and electron density, respectively.
A $14 \times 14 \times 14$ ($8 \times 8 \times 8$) MP special $k$-point set was used for the fcc (hcp) phase.
The optimized lattice constant and cohesive energy for fcc argon are summarized in Table \ref{table:Ar-fcc}, along with CCSD(T) and experimental values.
The lattice constant obtained using LDA (PBE) is underestimated (overestimated),whereas those with vdW-DFs are improved except for vdW-DF2$^{\rm C09x}$, and most of the vdW-DFs improve cohesive energy.
Our results for fcc argon are in good agreement with the recent results by Tran and Hutter.\cite{Tran2013}

The results for hcp argon are summarized in Table \ref{table:Ar-hcp}. 
The difference between the cohesive energies for the fcc and hcp phases is very small (at most 2 meV/atom) and these phases are almost degenerated.
The lattice parameters for the hcp structure satisfy the ideal ratio $a_{\rm fcc} \sim \sqrt{2} a_{\rm hcp}$, $c_{\rm hcp}/a_{\rm hcp} = 1.633$, where $a_{\rm fcc}$, $a_{\rm hcp}$, and $c_{\rm hcp}$ are the lattice constant for the fcc structure, the in-plane lattice constant for the hcp structure, and the out-of-plane-lattice constant for the hcp structure, respectively.
The deviations of $a_{\rm fcc}$ from the ideal value are 0.15 \AA~ for PBE, 0.36 \AA~for vdW-DF2$^{\rm C09x}$, and 0.03 \AA~for other functionals.
Our result indicates that these two phases are almost identical and cannot reproduce the experimentally observed stable fcc phase at ambient pressure.
This apparent discrepancy may be solved by introducing an entropic effect from phonons.
Indeed, Ishikawa {\it et al}.\cite{Ishikawa2014} performed lattice dynamics calculations and successfully predicted a stable fcc argon by taking into account the entropic contributions. 
%
\subsection{Graphite\label{secsec:gr}}

Graphite consists of the two-dimensional carbon allotrope, graphene, and graphene layers bound in the out-of-plane direction with a vdW interaction.
Graphite has been studied extensively both experimentally and theoretically, and is often used to assess the accuracy of a new method for weak interactions, as benchmark calculations with a highly accurate electronic structure method such as QMC\cite{Spanu2009} and ACFDT-RPA.\cite{Lebegue2010}

In our calculation, the Brillouin zone was sampled using an $8 \times 8 \times 4$ MP $k$-point set, and plane wave cutoffs of 80 and 480 Ry were used for wave functions and electron density, respectively.
We fully optimized the lattice parameters according to the calculated internal pressure, and binding energy was determined from the difference in total energy at the equilibrium and that at an interlayer distance larger than 13 \AA.
In Table~\ref{table:Graphite}, optimized lattice parameters and binding energies obtained using different exchange-correlation functionals are summarized.
\begin{table}[tb]
  \begin{center}
    \caption{In-plane lattice constant ($a$), out-of-plane lattice constant ($c$), equilibrium volume ($V_{0}$), and the binding energy $ \Delta E_{\rm b}$ for graphite. The numbers in parentheses were obtained using the experimental value.}
    \label{table:Graphite}
    \begin{tabular}{lcccc}  \hline \hline
      Functional           & $a$      & $c$      & $V_0$          & $\Delta E_{\rm b}$ \\
                           & (\AA)    & (\AA)    & (\AA$^3$/atom) & (meV/atom)             \\ \hline
      LDA                  & 2.45     & 6.69     & 8.71           & 24                   \\
      PBE                  & 2.47     & 8.76     & 11.61          & 2                   \\
      vdW-DF               & 2.49     & 7.18     & 9.61           & 55                   \\
      vdW-DF2              & 2.48     & 7.05     & 9.41           & 53                   \\
      vdW-DF$^{\rm C09x}$  & 2.47     & 6.50     & 8.60           & 76                   \\
      vdW-DF2$^{\rm C09x}$ & 2.47     & 6.58     & 8.71           & 56                   \\
      rVV10                & 2.48     & 6.76     & 8.93           & 68                   \\
      rVV10$^{a}$          & 2.46     & 6.72     & 8.80           & 39 \\
      VV10$^{b}$           & (2.46)   & 6.777    & (8.88)         & 71 \\
      vdW-DF-cx$^{c}$      & 2.46     & 6.43     & 8.54           & 66 \\
      QMC$^{d}$            & (2.46)   & 6.852    & (8.98)         & 56$\pm$5 \\
      ACFDT-RPA$^{e}$      & (2.46)   & 6.68     & (8.75)         & 48               \\
      Expt.                & 2.46$^f$ & 6.70$^f$ & 8.80$^f$       & 52 $\pm 5$$^g$         \\ \hline \hline
    \end{tabular}
  \end{center} 
  \begin{center}
    $^a$ Ref.~\onlinecite{Sabatini2013}. $^b$ Ref. \onlinecite{Bjoerkman2012}. $^c$ Ref. \onlinecite{Berland2014}. $^d$ Ref.~\onlinecite{Spanu2009}. $^e$ Ref. \onlinecite{Lebegue2010}. $^f$ Ref. \onlinecite{Baskin1955}. $^g$ Ref. \onlinecite{Zacharia2004}. 
  \end{center}
\end{table}

In accordance with the literature, LDA gives lattice parameters, which are in good agreement with the experimental results, whereas GGA yields a much larger out-of-plane lattice parameter ($c$) and negligible binding energy, suggesting that the latter cannot predict the binding of graphite.
The binding energy obtained with LDA is smaller than that obtained from the experiment and by a highly accurate theoretical method.
In general, our vdW-DF results are in good agreement with those of previous studies, that employed different vdW-DFs.\cite{Cooper2010,Lee2010,Hamada2010, Mapasha2012,Graziano2012, Berland2014}
vdW-DF and vdW-DF2 predict the binding energy ($\Delta E_{\rm b}$), in good agreement with the experimental results.
They also improve the description of structural properties, but the equilibrium volumes ($V_{0}$) are overestimated.
$V_{0}$ is improved using vdW-DF$^{\rm C09x}$ and rVV10, but they overestimate $\Delta E_{\rm b}$.
We note that our $\Delta E_{\rm b}$ obtained using rVV10 is almost twice as large as that reported by the original authors.\cite{Sabatini2013}
At present, the origin of the discrepancy is yet to be clarified, but similar larger values ($\sim$70 meV/atom) were obtained using different implementations \cite{JuNoLo,Lazic2010,Callsen2012} of rVV10 and different potentials.\cite{Callsen2014}
On the other hand, vdW-DF2$^{\rm C09x}$ predicts that lattice parameters and $\Delta E_{\rm b}$ are in good agreement with the experimental results, in line with a previous study.\cite{Hamada2010}

\subsection{Trigonal selenium \label{secsec:se}}

\begin{figure}
  \begin{center}
    \includegraphics[width=6cm, angle=359.5]{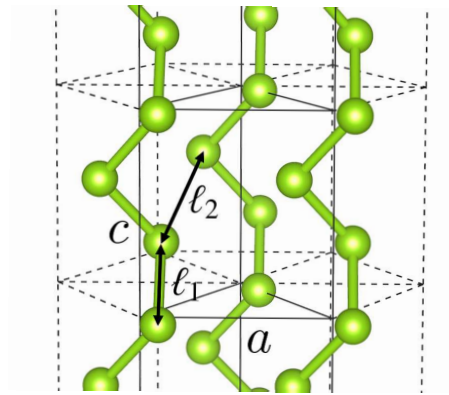}
    \caption{(Color online) Trigonal structure of selenium. \label{fig:Se-Trigonal}}
    \label{fig:Se-trigonal}
  \end{center}
\end{figure}

\begin{table*}[tb]
  \caption{Optimized lattice parameters ($a$ and $c$), equilibrium volume ($V_{0}$), internal parameter ($x$), shortest Se-Se distance in a chain ($\ell_{1}$), shortest Se-Se distance in different chains ($\ell_{2}$), and binding energy ($\Delta E_{\rm b}$) for trigonal selenium.}
  \label{table:Se-hcp}
  \begin{center}
    \begin{tabular}{lccccccc}
      \hline \hline
      Functional           & $a$       & $c$       & $V_0$          & $x$       & $\ell_{1}$ & $\ell_{2}$ & $\Delta E_{\rm b}$ \\
                           & (\AA)     & (\AA)     & (\AA$^3$/atom) &           & (\AA)      & (\AA)      & (meV/atom)  \\ \hline
      LDA                  & 3.89      & 5.04      & 21.94          & 0.2540    & 2.40       & 3.06       & 395 \\
      PBE                  & 4.51      & 4.97      & 29.16          & 0.2148    & 2.36       & 3.58       & 55 \\
      vdW-DF               & 4.70      & 5.04      & 32.15          & 0.2091    & 2.39       & 3.74       & 189 \\
      vdW-DF2              & 4.57      & 5.12      & 30.81          & 0.2177    & 2.42       & 3.62       & 228 \\
      vdW-DF$^{\rm C09x}$  & 3.96      & 5.11      & 23.08          & 0.2521    & 2.43       & 3.11       & 429 \\
      vdW-DF2$^{\rm C09x}$ & 3.99      & 5.10      & 23.42          & 0.2497    & 2.42       & 3.14       & 330 \\
      rVV10                & 4.24      & 5.11      & 26.54          & 0.2348    & 2.42       & 3.35       & 320 \\
      Expt.                 & 4.366$^a$ & 4.954$^a$ & 27.261$^a$     & 0.225$^b$ & 2.373$^a$  & 3.436$^a$  & 
      \\ \hline \hline
    \end{tabular}
  \end{center}
  \begin{center}
    $^a$ Ref. \onlinecite{Swanson1955}. $^b$ Ref. \onlinecite{Cherin1967}.
  \end{center}
\end{table*}
Trigonal selenium is formed by a bundle of one-dimensional covalently bonded atomic chiral  chains.
As already shown in previous works,\cite{DalCorso1994, Kresse1994, Bucko2010} the gradient correction to LDA improves the description of selenium, but is still less satisfactory, presumably because of the lack of vdW interaction between the chiral chains in GGA.
Bu\v{c}ko {\it et al}.\cite{Bucko2010} demonstrated using the semi-empirical dispersion correction that the structural parameters are significantly improved, and the agreement with the experimental results becomes much better.
Here, we use  vdW-DF to address the importance of vdW interaction.
The trigonal selenium  belongs either to the $P3_{1}21$ or $P3_{2}21$ space group and has three atoms per unit cell (see Fig. \ref{fig:Se-trigonal}).
The atomic position in a cell can be described using the internal parameter $x$, which scales the distance from the screw axis to the atomic position.
The structure of the trigonal selenium is also characterized by the shortest Se-Se distance in the same chain ($\ell_{1}$) and the shortest Se-Se distance in different chains ($\ell_{2}$).
We used an $8 \times 8 \times 4$ MP $k$-point set and plane wave cutoffs of 60 and 500 Ry for wave functions and electron density, respectively.
Binding energy was calculated from the difference between the total energies of a solid in equilibrium and the isolated chain.
Calculated structural parameters and binding energy are summarized in Table~\ref{table:Se-hcp}.
In general, the lattice constant along the chiral chain ($c$) is overestimated by all the functionals used in this study, whereas the accuracy for the lattice constant $a$ varies.
LDA significantly underestimates $a$, whereas PBE significantly improves the description of $a$, in good agreement with previous studies.
Both vdW-DF and vdW-DF2 overestimate $a$ and $c$, whereas the use of the C09 exchange (vdW-DF$^{\rm C09x}$ and vdW-DF2$^{\rm C09x}$) leads to the underestimation of $a$.
On the other hand, although $c$ is slightly overestimated, rVV10 provides a balanced description of both the lattice constants.
It is found from our results that there is a correlation between calculated $a$ and $c$:
if $a$ is overestimated, the error in $c$ tends to be smaller.
Thus, to obtain accurate structural parameters for the trigonal selenium, it is very important to describe both covalent and vdW interactions accurately, suggesting that this material can be a good benchmark system for a vdW inclusive functional.
%

\subsection{Dry ice \label{secsec:co2}}

\begin{figure}
  \begin{center}
    \includegraphics[width=5cm]{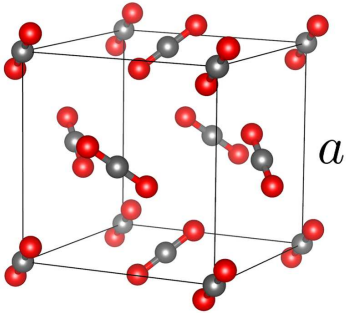}
    \caption{(Color online) Cubic $Pa3$ structure of dry ice (solid CO$_{2}$).}
    \label{fig:CO2-I-Structure}
  \end{center}
\end{figure}

\begin{table}[tb]
  \begin{center}
    \caption{Equilibrium lattice parameter ($a$), volume ($V_{0}$), C-O bond length ($\ell_{\rm b}$), and binding energy ($\Delta E_{\rm b}$) for dry ice.}
    \begin{tabular}{lcccc}
      \hline \hline
      Functional             & $a$         & $V_0$                         & $\ell_{\rm b}$ & $\Delta E_{\rm b}$   \\
                             & (\AA)       & (${\rm \AA ^3/{\rm CO}_{2}}$) & (${\rm \AA}$)  & (meV/CO$_{2}$) \\  \hline
      LDA                    & 5.28        & 36.88                         & 1.165          & 370       \\
      PBE                    & 6.04        & 55.11                         & 1.175          & 103       \\
      vdW-DF                 & 5.77        & 47.94                         & 1.180          & 364       \\
      vdW-DF2                & 5.61        & 44.04                         & 1.178          & 346       \\
      vdW-${\rm DF^{C09x}}$  & 5.53        & 42.17                         & 1.176          & 339       \\
      vdW-${\rm DF2^{C09x}}$ & 5.79        & 48.42                         & 1.176          & 155       \\
      rVV10                  & 5.51        & 41.72                         & 1.179          & 328       \\
      MP2$^{a}$ & 5.46       & 40.69                         & 1.17           & 290 \\
      Expt.                  & 5.62~$^b$  & 44.38$^b$                    & 1.155$^b$      & 288$^{c}$   \\
      \hline \hline
    \end{tabular}
    \label{table:Solid-CO2-I}
  \end{center}
  \begin{center}
    $^a$ Ref. \onlinecite{Sode2013}. $^b$ Ref. \onlinecite{Simon1980}. $^c$ Ref. \onlinecite{Otero-de-la-Roza2012}.
  \end{center}
\end{table}

%
%

The solid form of carbon dioxide is called dry ice.
Carbon dioxide has no dipole moment, and thus attractive vdW forces play an important role in the condensation.
We chose this material as a representative application of our vdW-DF implementation to  non magnetic molecular crystals.
Crystalline dry ice has a cubic symmetry with the space group of $Pa3$ (see Fig.~\ref{fig:CO2-I-Structure} for the structure).
Plane wave cutoffs of 160 and 960 Ry were used for wave functions and electron densities, respectively.
An $8 \times 8 \times 8$ MP $k$-point set was used for Brillouin zone sampling. 
Optimized structural parameters and binding energy for the dry ice are given in Table \ref{table:Solid-CO2-I}, along with the MP2\cite{Sode2013} and experimental (150K)\cite{Simon1980} results.
Our PBE equilibrium volume is in good agreement with the theoretical value obtained by Bonev {\it et al.}\cite{Bonev2003} (52.9 ${\rm \AA ^3/{\rm CO}_{2}}$) using the same functional.
On the other hand, both the equilibrium volume and binding energy obtained using vdW-DFs are in better agreement with the highly accurate MP2 and experimental (150K) results, suggesting the improvement over LDA and PBE.
The maximum deviation of the equilibrium volume obtained using vdW-DF is 9.1\% compared with that obtained in the low-temperature experiment.
The above results suggest that PBE overestimates the equilibrium volume because of the lack of dispersion forces, and a more accurate description of the structure and energetics of dry ice is made possible by considering the dispersion forces with vdW-DF.
Regarding the severe underestimation of the binding energy using vdW-DF2$^{\rm C09x}$, it has been found\cite{Hamada2013} from a systematic assessment using the S22 dataset that although it predicts reasonable intermolecular separation (distance), the functional severely underestimates the binding energy, because the C09 exchange is too repulsive at a relatively large density gradient relevant to the intermolecular region.
This result implies that vdW-DF2$^{\rm C09x}$ inaccurately describes the intermolecular vdW interaction.
%

\subsection{Solid oxygen in the $\alpha$ phase \label{secsec:o2}}

\begin{figure}
  \begin{center}
    \includegraphics[width=7cm]{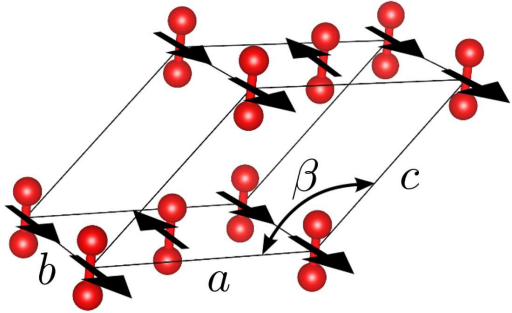}
    \caption{(Color online) Crystal structure of solid oxygen in $\alpha$ phase. 
                The arrows on molecules represent the magnetic moment.}
    \label{fig:o2}
  \end{center}
\end{figure}

\begin{table*}[tb]
  \begin{center}
    \caption{Optimized lattice parameters ($a$, $b$, $c$, and $\beta$), nearest-neighbor distance ($\ell_{\rm mol}\equiv \sqrt{a^2+b^2}/2$), equilibrium volume ($V_{0}$), bond length ($\ell_{\rm b}$), magnetic moment ($M_a$) on oxygen atom, and binding energy of molecule ($\Delta E_{b}$). Experimental values are shown for comparison.}
\begin{tabular}{lccccccccc} 
\hline \hline
Functional              & $a$        & $b$        & $c$        & $\beta$    & $\ell_{\rm mol}$ & $V_0$       & $\ell_b$  & $M_a$           & $\Delta E_b$ \\
                        & (\AA)      & (\AA)      & (\AA)      & (deg)      & (\AA)            & (\AA$^{3}$) & (\AA)     & ($\mu_{\rm B}$) & (meV)  \\ \hline
LDA                     & 3.29       & 3.28       & 4.05       & 113.9      & 2.32             & 19.93       & 1.202     & 0.30            & 552        \\
PBE                     & 4.59       & 3.93       & 5.05       & 122.1      & 3.02             & 38.54       & 1.218     & 0.66            & 41         \\
vdW-DF                  & 4.68       & 3.68       & 4.70       & 125.2      & 2.98             & 33.05       & 1.231     & 0.66            & 221       \\
vdW-DF-GC               & 4.94       & 3.57       & 4.91       & 128.4      & 3.05             & 33.85       & 1.231     & 0.66            & 213        \\
vdW-DF2                 & 3.83       & 3.85       & 4.29       & 118.6      & 2.72             & 27.75       & 1.235     & 0.60            & 225        \\
vdW-DF2-GC              & 3.91       & 3.88       & 4.30       & 119.0      & 2.76             & 28.56       & 1.235     & 0.62            & 209        \\
vdW-DF$^{\rm C09x}$     & 3.52       & 3.46       & 4.15       & 114.8      & 2.47             & 22.90       & 1.215     & 0.47            & 285        \\
vdW-DF$^{\rm C09x}$-GC  & 3.59       & 3.47       & 4.17       & 115.4      & 2.50             & 23.51       & 1.215     & 0.50            & 255         \\
vdW-DF2$^{\rm C09x}$    & 3.66       & 3.56       & 4.34       & 115.1      & 2.55             & 25.51       & 1.216     & 0.52            & 95          \\
vdW-DF2$^{\rm C09x}$-GC & 3.79       & 3.62       & 4.36       & 115.8      & 2.62             & 26.92       & 1.217     & 0.57            & 71          \\
rVV10                   & 3.71       & 3.62       & 4.18       & 117.2      & 2.59             & 24.94       & 1.225     & 0.56            & 240        \\
Expt.$^a$                   & 5.403 & 3.429 & 5.086 & 132.3 & 3.200       & 34.85  & 1.29 &                 & \\
\hline \hline
\end{tabular}
\label{table:Solid-Oxygen-alpha}
\end{center}
\begin{center}
    $^a$ Ref. \onlinecite{Meier1984}.
%
%
\end{center}
\end{table*}

\begin{table}[tb]
  \begin{center}
    \caption{Magnetic energy for $\alpha$-O$_{2}$ ($\Delta E^{\rm mag}$).}
\begin{tabular}{lc}
\hline \hline
Functional & $\Delta E^{\rm mag}$ \\
           & (meV) \\ \hline
PBE        & 95  \\
vdW-DF     & 118 \\
vdW-DF-GC  & 87  \\
vdW-DF2    & 304 \\
vdW-DF2-GC & 251 \\ \hline \hline
\end{tabular}
\label{table:emag}
\end{center}
\end{table}
$\alpha$-O$_{2}$ has an antiferromagnetic ground state with the crystal structure of the $C2/m$ space group, as shown in Fig. \ref{fig:o2}.
There are four lattice parameters, $a$, $b$, $c$, and $\beta$, and the internal parameters of  $\ell_{\rm b}$ (bond length of O$_{2}$ molecule) and $\theta$ (tilted angle of molecular axis).
The molecular axis is almost perpendicular to the $ab$-plane and slightly tilted within the $ac$-plane. 
The angle $\theta$ was reported to be $\sim$3$^\circ$  in the experiment.\cite{Meier1984}.
In the calculation,  we used plane wave cutoffs of 160 and 960 Ry for wave functions and electron density, respectively, and an $8 \times 8 \times 8$ MP $k$-point set was used for the Brillouin zone sampling.
During the structural optimization, the molecular axis was fixed in the direction perpendicular to the $ab$-plane, but the residual forces on the atom were small, typically less than 0.001 eV/\AA.
In Table \ref{table:Solid-Oxygen-alpha}, we report the structural parameters and binding energies obtained using different functionals.
As expected, LDA severely underestimates the equilibrium volume by 42.8\%, whereas PBE overestimates it, but the error is marginal (10.6\%).
However, this unexpectedly small error is due to the cancellation of the errors in $a$ ($-$15.0\%) and $b$ (14.6\%).
The error in $c$ is surprisingly small, but the trend in the lattice parameters estimated using PBE is not systematic.
On the other hand, all the vdW-DFs underestimate $a$ and $c$, whereas $b$ is overestimated, and as a result, the equilibrium volumes are consistently underestimated.
The angle $\beta$ is also consistently underestimated.
Among vdW-DFs, the original vdW-DF proposed by Dion {\it et al}. predicts the most accurate structural parameters, and inclusion of GC further improves the structural parameters, which are in good agreement with the experimental results, suggesting the importance of the spin-polarization-dependent gradient correction of the local correlation.

The nearest-neighbor distance between O$_{2}$ molecules ($\ell_{\rm mol} \equiv \sqrt{a^{2}+b^{2}}/2$) is a very important quantity in $\alpha$-O$_{2}$, as it strongly correlates with the antiferromagnetic interaction between O$_{2}$ molecules.
We found that $\ell_{\rm mol}$ is underestimated by LDA and PBE, but is increased using vdW-DF-GC to a distance of 3.05 \AA~which is close to the experimental value (3.20 \AA).
We also note that there is a correlation between $\beta$ and $c$ estimated using vdW-DF, i.e., the smaller the $c$, the smaller the $\beta$.
As discussed in previous works,\cite{Obata2013,Santoro2001} the distance between magnetic molecules is determined by a subtle balance between magnetic and vdW interactions.
In the solid state, not only the balance between magnetic and vdW interactions within the $ab$-plane, but also that between $ab$-planes (out-of-plane direction) plays a decisive role:
a complicated interplay among the antiferromagnetic, ferromagnetic, and vdW interactions in three dimensions determines the structure of $\alpha$-O$_{2}$.
The molecules in the $ab$-plane can be considered to form a triangular configuration. 
In the experimental structure, this configuration is very similar to an equilateral triangle, where the antiferromagnetic interaction between the neighboring molecules could destabilize the magnetic structure or distort the triangle to reduce the magnetic frustration.
On the basis of this consideration and the knowledge acquired in a previous study\cite{Obata2013}, the underestimation of the nearest-neighbor distance ($\ell_{\rm mol}$) using vdW-DF may be ascribed to the overestimation of the antiferromagnetic interaction between molecules with respect to the ferromagnetic one.
The structural properties depend strongly on the exchange energy functional used, as shown in Table \ref{table:Solid-Oxygen-alpha}.
Such a behavior has also been shown in the previous study on the pair of oxygen molecules\cite{Obata2013}.
The revPBE exchange functional in the original vdW-DF gives a more repulsive intermolecular potential than the other exchange functionals.
The other vdW-DFs consisting of different exchange functionals (C09 and PW86R) underestimate $\ell_{\rm mol}$, indicating that these exchange functionals enhance the overstabilization of the antiferromagnetic state.
Thus, to improve the description of $\alpha$-O$_{2}$, it is necessary to develop more accurate exchange and correlation functionals that predict a balanced description of antiferromagnetic and ferromagnetic interactions.
The magnetic interaction $J$ is proportional to $-t^{2}/\Delta E$, where $t$ and $\Delta E$ are the transfer integral and the energy gap between the spin-up and spin-down states immediately above and below the Fermi level, respectively.
Because GGA is known to underestimate the energy gap for insulating and molecular systems, the antiferromagnetic states are overstabilized as a result of the underestimation of $\Delta E$ (overestimation of $\vert t \vert$ as well).
To capture a trend in the functionals on the magnetic interaction, we calculated magnetic energy, defined as the energy difference between the ferromagnetic (F) and antiferromagnetic(AF) states at the same AF crystal structure ($\Delta E^{\rm mag}=E^{\rm F}-E^{\rm AF}$).
The results obtained using PBE, vdW-DF, and vdW-DF2 (with and without GC) are summarized in Table~\ref{table:emag}.
It is found that vdW-DF, which predicts more accurate lattice parameters (larger $\ell_{\rm mol}$) than vdW-DF2, yields a smaller $\Delta E^{\rm mag}$, suggesting that the intermolecular distance plays an important role in the magnetic interaction.
Inclusion of GC leads to the enlargement of the nearest O$_{2}$ distance ($\ell_{\rm mol}$) and reduction of $\Delta E^{\rm mag}$, because GC destabilizes the antiferromagnetic states, which works as a repulsive force between the nearest O$_{2}$ molecules\cite{Obata2013}.
%
%
\section{Summary \label{sec:summary}} 
%

In this work, we have studied solid oxygen in the $\alpha$-phase with the antiferromagnetic configuration, using several variants of vdW-DF and the correction scheme for magnetic systems proposed in our previous study.
To test the vdW-DF energy, force, and stress implemented in our program code, we performed the calculations for solid argon in the fcc and hcp phases, graphite, trigonal selenium, and  dry ice, in which the vdW interaction is considered to be important.
We have found that vdW-DF shows an overall improvement in the structure and energy of these materials over LDA and GGA, in good agreement with previous studies.
In the case of antiferromagnetic solid oxygen, we have found that vdW-DFs consistently underestimate the lattice parameters, and the original vdW-DF proposed by Dion {\it et al}. with our spin-polarization-dependent gradient correction scheme (vdW-DF-GC) provides the most accurate structural parameters among other functionals.
We point out the competing nature of ferromagnetic, antiferromagnetic, and vdW interactions in a solid oxygen, and for a more accurate prediction of the structural properties of solid oxygen, a balanced description of these interactions is decisively important.
Thus, we suggest that solid oxygen can be a critical test material for an electronic structure method, which can account for magnetic and vdW interactions accurately.
Applications of the present vdW-DF scheme for the magnetic systems to the molecular magnet or metalorganic molecular systems are highly anticipated, which promote research on molecular spintronics.

%
\begin{acknowledgments}
%

The computation in this work was performed using the facilities of the Supercomputer Center, Institute for Solid State Physics, University of Tokyo, and the facilities of the Research Center for Computational Science, National Institutes of Natural Sciences, Okazaki, Japan.
This work was partly supported by Grants-in-Aid for Scientific Research from JSPS/MEXT (Grant Nos. 22104012 and 26400312),  the Strategic Programs for Innovative Research (SPIRE), MEXT, Japan, the Computational Materials Science Initiative (CMSI), Japan, and the World Premier International Research Center Initiative (WPI) for Materials Nanoarchitectonics, MEXT, Japan.

\end{acknowledgments}

%
%
\begin{appendix}
%
%
\section{rVV10 with the Wu-Gygi implementation \label{appendix2}}
%
The rVV10 functional has been proposed\cite{Sabatini2013} for implementing the VV10 functional\cite{Vydrov2010} with the efficient RPS algorithm.\cite{Roman-Perez2009}
We note that, very recently, the original VV10 has been implemented using the WG scheme by Corsetti {\it et al}.\cite{Corsetti2013}
Here, we briefly describe our implementation of rVV10 with the algorithm proposed by WG.\cite{Wu2012}
The rVV10 nonlocal correlation functional is given by
\begin{align}
  E_{\rm c}^{\rm nl} =& \frac{1}{2} \iint \frac{n({\bf r}_1)}{k^{3/2}({\bf r}_1)}
  \phi^{\rm rVV10}(\tilde{q}_1,\tilde{q}_2,r_{12}) \frac{n({\bf r}_2)}{k^{3/2}({\bf r}_2)} d{\bf r}_1 d{\bf r}_2,
  \label{eq:App0-1}
\end{align}
where $\phi^{\rm rVV10}(\tilde{q}_1,\tilde{q}_2,r_{12})$ is the rVV10 nonlocal correlation kernel,\cite{Sabatini2013}  $\tilde{q}_1=\tilde{q}({\bf r}_1)$, $\tilde{q}_2=\tilde{q}({\bf r}_2)$, and $k(n({\bf r}))=3\pi b(n/9\pi)^{1/6}/2$.
$\tilde{q}_{i}$ is defined as $\tilde{q}({\bf r}_{i})=\omega_0(n({\bf r}_{i}),\vert \nabla n ({\bf r}_{i}) \vert)/k(n({\bf r}_{i}))$ with $\omega_0$ defined in Ref.~\onlinecite{Vydrov2010}.
The parameter $b$ is an adjustable parameter, which is determined by fitting on a training set (see Refs.~\onlinecite{Vydrov2010} and \onlinecite{Sabatini2013} for details).
By expanding the nonlocal correlation kernel in the same manner as Eq.~(\ref{eq:phi}), the nonlocal correlation energy is written as
\begin{align}
  E_{\rm c}^{\rm nl} =& \frac{\Omega}{2} \sum _{\bf G} \sum_{\alpha\beta} \tilde{\eta}^*_{\alpha} ({\bf G})\tilde{\eta}_{\rm \beta}({\bf G}) \phi^{\rm rVV10}_{\alpha\beta}(G),
  \label{eq:App0-2}
\end{align}
where $\tilde{\eta}_{\alpha}({\bf G})$ is the Fourier component of $\tilde{\eta}_{\alpha} ({\bf r})$, which is defined by $\tilde{\eta}_{\alpha} ({\bf r}) = {{q}_{\alpha}n({\bf r})p_{\alpha}(\tilde{q}({\bf r}))}/{k^{3/2}({\bf r}) \tilde{q}({\bf r})}$, and the Fourier transform of the nonlocal correlation kernel $\phi^{\rm rVV10}$ is given by
\begin{align}
  \phi^{\rm rVV10}_{\alpha\beta}(G) =&  \frac{1}{{q_\alpha}^{3/2}} F^{\rm rVV10}_{\alpha\beta} \left( \frac{G}{\sqrt{q_\alpha}}\right), \\
   F^{\rm rVV10}_{\alpha\beta}(k) =& \frac{4\pi}{k} \int_0^{\infty} u \sin(ku) \phi^{\rm rVV10} \left( u,\sqrt{\frac{q_\beta}{q_\alpha}}u \right ) du .
  \label{eq:App0-3}
\end{align}
The nonlocal correlation potential is calculated similarly to Eq.~(\ref{eq:v-ecnl}).
%
%
\section{Fourier component of the kernel function $\phi_{\alpha\beta}(G)$\label{appendix1}}
\begin{figure}
  \begin{center}
    \includegraphics[width=8.0cm]{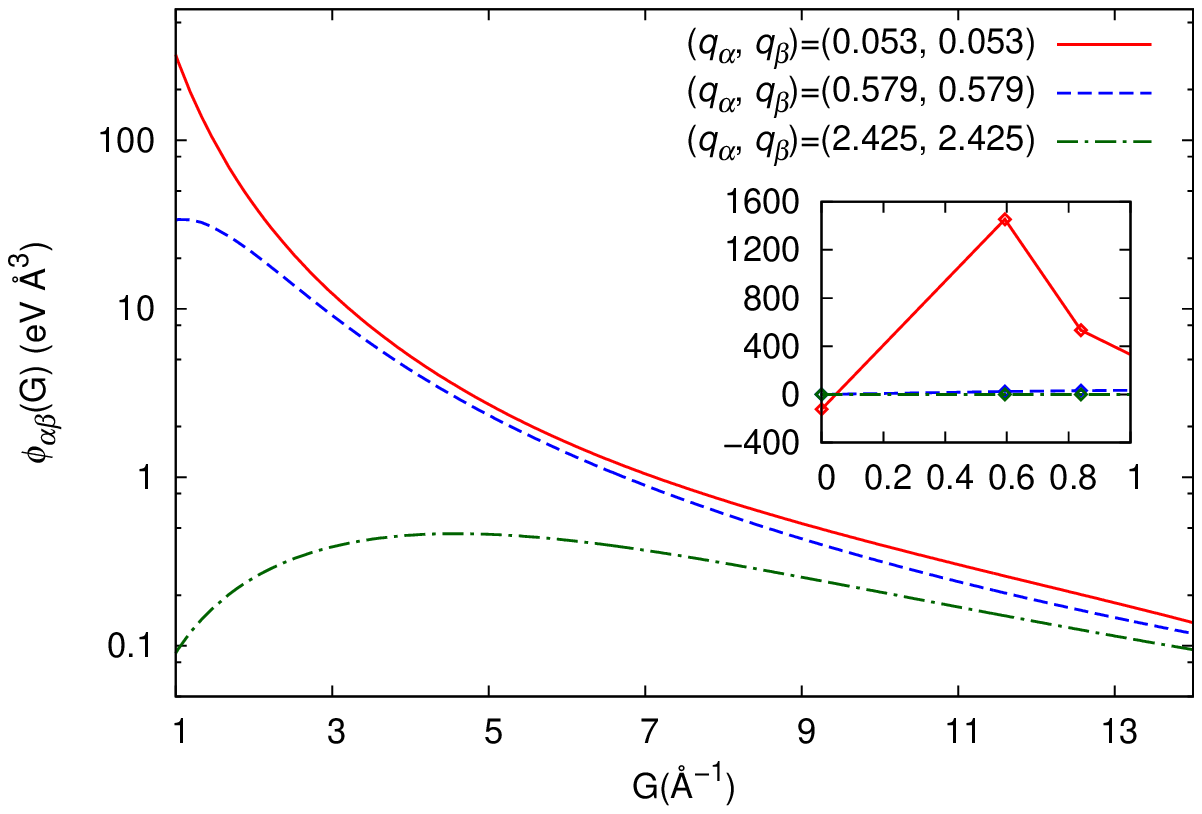}
    \includegraphics[width=8.0cm]{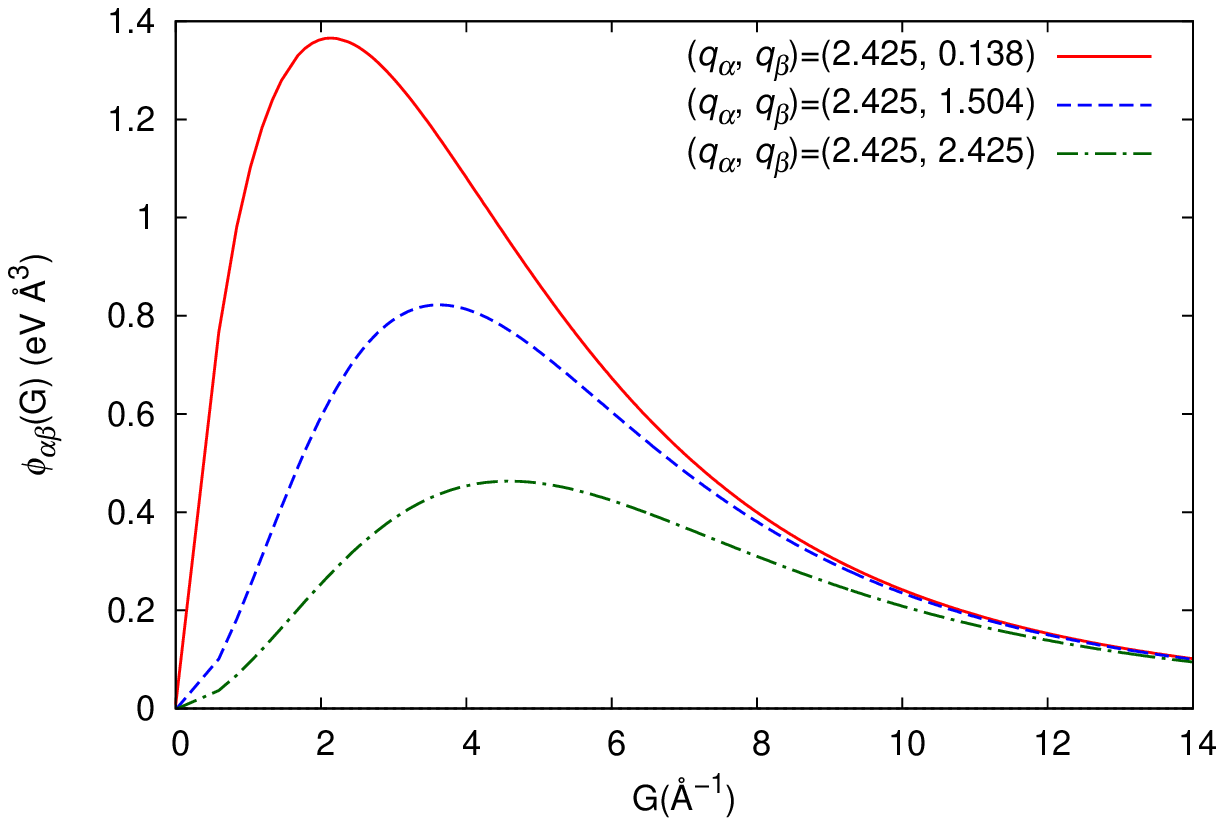}
    \caption{(Color online) Fourier component of the kernel function $\phi_{\alpha\beta}(G)$ for several sets of $q_\alpha$ and $q_\beta$. Diagonal (upper panel) and off-diagonal (lower panel) components are shown.}
    \label{fig:phiabg}
  \end{center}
\end{figure}

In the present implementation, the Fourier components of the kernel function $\phi_{\alpha \beta}(G)$ are calculated from the tabulated data for $\phi(d_1,d_2)$ ($d_1\phi(d_1,d_2)$ in the case of WG implementation) using Eq.~(\ref{eq:phi-alpha-beta-g}).
The typical functions of $\phi_{\alpha\beta}(G)$ are presented in Fig. \ref{fig:phiabg}.
For the small $q_{\alpha}$'s,  $\phi_{\alpha \alpha}(G)$ deviates slightly from zero at $G=0$  in the present computation, which may cause a numerical error.
However, the error in the absolute value of $E_{\rm c}^{\rm nl}$ associated with the deviation is typically $\sim$0.5 meV in fcc argon and negligible in the estimation of cohesive energy.
%
\section{Distribution of $q_{0}$ function \label{appendix3}}
%
%
%
\begin{figure}
  \begin{center}
    \includegraphics[width=8.5cm]{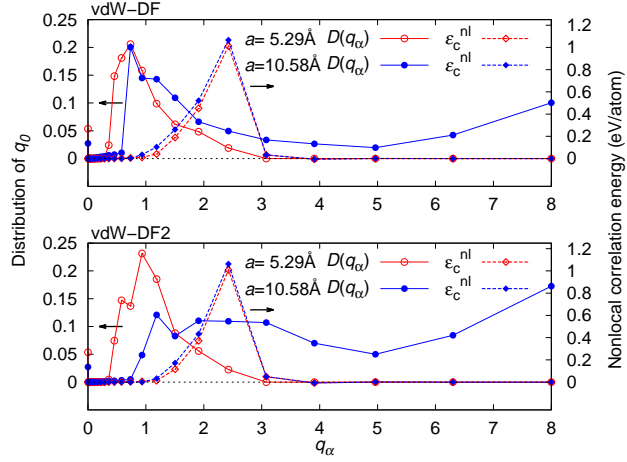}
    \caption{(Color online) Distributions of $q_0$ function (solid curves) and $\varepsilon_{\rm c}^{\rm nl}(\alpha)$ (dotted curves) for fcc argon calculated using vdW-DF (upper panel) and vdW-DF2 (lower panel). The red open and blue filled symbols are for the lattice parameters ($a$) of 5.29 and 10.58 \AA, respectively.}
    \label{fig:Ar-fcc-qmesh}
  \end{center}
\end{figure}

To expand Eq.~(\ref{eq:phi}), an appropriate cutoff $q_{c}$ and mesh for the $q_0$ function ($q_{\alpha}$) are needed.
To determine $q_c$ and $q_{\alpha}$, we investigated the distribution of the $q_{0}$ function by sampling the values on each FFT grid point.
The distribution $D(q_0)$ was calculated as a histogram by counting the number of samples while satisfying $q_\alpha \le q_0({\bf r}) < q_{\alpha+1}$.
In Fig. \ref{fig:Ar-fcc-qmesh}, $D(q_0)$'s for the vdW-DF and vdW-DF2 are plotted for different lattice parameters. 
Furthermore, we analyzed the contribution of the nonlocal correlation energy from each $q_{\alpha}$ value, by defining
\begin{align}
 E_{\rm c}^{\rm nl} &= \sum_{\alpha} \varepsilon_{\rm c}^{\rm nl}(\alpha),
\nonumber \\
 \varepsilon_{\rm c}^{\rm nl}(\alpha)  &=
  \frac{ \Omega}{ 2} \sum_{ {\bf G}}^{ } \left( |\eta_\alpha ({\bf G})|^2  
 \phi_{\alpha\alpha}(G) 
 + 2\sum_{ \beta (< \alpha) }^{} {\rm Re} (\eta _{\alpha}^* ({\bf G}) 
    \eta_\beta ({\bf G}) ) \phi_{\alpha\beta}(G) \right),
 \label{eq:varepsilon} 
\end{align} 
where $\varepsilon_{\rm c}^{\rm nl}(\alpha)$ is a positive value.
$\varepsilon_{\rm c}^{\rm nl}(\alpha)$'s for fcc argon calculated using vdW-DF and vdW-DF2 are shown in Fig. \ref{fig:Ar-fcc-qmesh}.
There is almost no contribution of the nonlocal correlation energy at the smallest $q_\alpha$, although the samples of $D(q_\alpha)$ exist there. 
Note that the samples at the smallest $q_\alpha$ come from diluted electron densities.
In the case of $a=5.29$~\AA, $\varepsilon_{\rm c}^{\rm nl}(\alpha)$'s have a peak at $q=2.4$ and are damped rapidly as $q_\alpha$ increases.
In the case of $a=10.58$~\AA, the shrunken electron cloud around atomic cores causes  the distributions of $D(q_\alpha)$ at larger $q_\alpha$'s($>5$).
However, there is no energy contribution from the $E_{\rm c}^{\rm nl}$ term.
This is presumably because the vdW interaction is both canceled out in the atom and fully damped at the atomic distances.
From the above result, it is confirmed that the $q_\alpha$ mesh defined in the present work covers the energy range in the system. 

\end{appendix}
%


\begin{thebibliography}{99}
\bibitem{Freiman2004}
Y. A. Freiman and H. J. Jodl,
Physics Reports {\bf 401}, 1 (2004).  

\bibitem{DeFotis1981}
G. C. DeFotis, 
Phys. Rev. B {\bf 23}, 4714 (1981).

\bibitem{Klotz2010}
S. Klotz, Th. Str{\" a}ssle, A. L. Cornelius, J. Philippe, and Th. Hansen, 
Phys. Rev. Lett. {\bf 104}, 115501 (2010). 

\bibitem{Meier1984}
R. J. Meier and R. B. Helmholdt, 
Phys. Rev. B {\bf 29}, 1387 (1984).

\bibitem{Etters1985}
R.~D.~Etters, K.~Kobashi, and J.~Belak,
Phys. Rev. B {\bf 32}, 4097 (1985).

\bibitem{Otani1998}
M. Otani, K. Yamaguchi, H. Miyagi, and N. Suzuki, 
J. Phys.: Condens. Matter {\bf 10}, 11603 (1998).

\bibitem{Kususe1999}
K. Kususe, Y. Hori, S. Suzuki, and K. Nakao, 
J. Phys. Soc. Jpn. {\bf 68}, 2692 (1999).

\bibitem{Nozawa2002}
K. Nozawa, N. Shima, and K. Makoshi, 
J. Phys. Soc. Jpn. {\bf 71}, 377 (2002).

\bibitem{Nozawa2008}
K. Nozawa, N. Shima, and K. Makoshi, 
J. Phys.: Condens. Matter {\bf 20}, 335219 (2008). 

\bibitem{Santoro2001}
M. Santoro, F. A. Gorelli, L. Ulivi, R. Bini, and H. J. Jodl, 
Phys. Rev. B {\bf 64}, 064428 (2001).

\bibitem{Grimme2006}
S.~Grimme,
J. Comput. Chem. {\bf 27}, 1787 (2006).

\bibitem{Grimme2010}
S. Grimme, J. Antony, S. Ehrlich and H. Krieg,  
J. Chem. Phys. {\bf 132}, 154104 (2010). 

\bibitem{Tkatchenko2009}
A.~Tkatchenko and M.~Scheffler,
Phys. Rev. Lett. {\bf 102}, 073005 (2009).

\bibitem{Tkatchenko2012}
A.~Tkatchenko, R.~A.~DiStasio, Jr., R.~Car, and M.~Scheffler,
Phys. Rev. Lett. {\bf 108}, 236402 (2012).

\bibitem{Silvestrelli2008}
P.~L.~Silvestrelli,
Phys. Rev. Lett. {\bf 100}, 053002 (2008).

\bibitem{Silvestrelli2013}
P.~L.~Silvestrelli,
J. Chem. Phys. {\bf 139}, 054106 (2013).

\bibitem{Rydberg2003}
H.~Rydberg, M.~Dion, N.~Jacobson, E.~Schr{\" o}der, P.~Hyldgaard, S.~I.~Simak, D.~C.~Langreth, and B.~I.~Lundqvist,
Phys. Rev. Lett. {\bf 91}, 126402 (2003).

\bibitem{Dion2004} 
M. Dion, H. Rydberg, E. Schr\"oder, D. C. Langreth, and B. I. Lundqvist, 
Phys. Rev. Lett. {\bf 92}, 246401 (2004) [Erratum  {\bf 95}, 109902 (2005)].

\bibitem{Vydrov2009}
O.~A.~Vydrov and T.~Van Voorhis,
Phys. Rev. Lett. {\bf 103}, 063004 (2009).

\bibitem{Klimes2010}
J. Klime\v{s}, D. R. Bowler, and A. Michaelides,
J. Phys.: Condens. Matter {\bf 22}, 022201 (2010).

\bibitem{Cooper2010}
V.~R.~Cooper,
Phys. Rev. B {\bf 81}, 161104(R) (2010).

\bibitem{Lee2010}
K.~Lee, \'{E}.~D.~ Murray, L.~Kong, B.~I.~Lundqvist, and D.~C.~Langreth, 
Phys. Rev. B {\bf 82}, 081101(R) (2010).

\bibitem{Vydrov2010}
O.~A.~Vydrov and T.~Van Voorhis,
J. Chem. Phys. {\bf 133}, 244103 (2010).

\bibitem{Klimes2011} 
J. Klime\v{s}, D. R. Bowler, and A. Michaelides,
Phys. Rev. B {\bf 83}, 195131 (2011).

\bibitem{Wellendorff2012}
J.~Wellendorff, K.~T.~Lundgaard, A. M{\o}gelh{\o}j, V.~Petzold, D.~D.~Landis, J.~N{\o}rskov, T. Bligaard, and K.~W.~Jacobsen,
Phys. Rev. B {\bf 85}, 235149 (2012).

\bibitem{Sabatini2013}
R. Sabatini, T. Gorni, and S. de Gironcoli, 
Phys. Rev. B {\bf 87}, 041108(R) (2013). 

\bibitem{Berland2014}
K.~Berland and P.~Hyldgaard,
Phys. Rev. B {\bf 89}, 035412 (2014).

\bibitem{Hamada2014}
I.~Hamada,
Phys. Rev. B {\bf 89}, 121103(R) (2014).

\bibitem{Obata2013}
M.~Obata, M.~Nakamura, I.~Hamda, and T.~Oda,
J. Phys. Soc. Jpn. {\bf 82}, 093701 (2013).

\bibitem{Roman-Perez2009} 
G. Rom\'{a}n-P\'{e}rez and J. M. Soler,
Phys. Rev. Lett. {\bf 103}, 096102 (2009).

\bibitem{Wu2012}
J. Wu and F. Gygi,
J. Chem. Phys. {\bf 136}, 224107 (2012). 

\bibitem{White1994}
J.~A.~White and D.~M.~Bird,
Phys. Rev. B {\bf 50}, 4954 (1994).

\bibitem{Sabatini2012}
R.~Sabatini, E.~K{\" u}\c{c}{\" u}kbenli, B.~Kolb, T.~Thonhauser, and S.~de Gironcoli,
J. Phys.: Condens. Matter {\bf 24}, 424209 (2012).

\bibitem{Perdew1996}
J. P. Perdew, K. Burke, and M. Ernzehof, 
Phys. Rev. Lett. {\bf 77}, 3865 (1996).

\bibitem{Car1985}
R.~Car and M.~Parrinello,
Phys. Rev. Lett. {\bf 55}, 2471 (1985).

\bibitem{Oda1998}
T.~Oda, A.~Pasquarello, and R.~Car,
Phys. Rev. Lett. {\bf 80}, 3622 (1998).

\bibitem{Oda2002}
T.~Oda,
J. Phys. Soc. Jpn. {\bf 71}, 519 (2002).

\bibitem{Vanderbilt1990}
D. Vanderbilt,
Phys. Rev. B {\bf 41}, 7892 (1990).

\bibitem{Laasonen1993}
K. Laasonen, A. Pasquarello, R. Car, C. Lee, and D. Vanderbilt, 
Phys. Rev. B {\bf 47}, 10142 (1993).

\bibitem{Perdew1986}
J.~P.~Perdew and Y.~Wang,
Phys. Rev. B {\bf 33}, 8800(R) (1986).

\bibitem{Murray2009}
\'{E}.~D.~Murray, K.~Lee, and D.~C.~Langreth,
J. Chem. Theory Comput. {\bf 5}, 2754 (2009).


\bibitem{Monkhorst1976}
H. J. Monkhorst and J. D. Pack,
Phys. Rev. B {\bf 13}, 5188 (1976).

\bibitem{Perdew1981}
J. P. Perdew and A. Zunger,
Phys. Rev. B {\bf 23}, 5048 (1981).

\bibitem{Zhang1998}
Y. Zhang and W. Yang,
Phys. Rev. Lett. {\bf 80}, 890 (1998).

\bibitem{Tran2013}
F.~Tran and J.~Hutter,
J. Chem. Phys. {\bf 138}, 204103 (2013).

\bibitem{Posciszewski2000}
K. Ro\'{s}ciszewski, B. Paulus, P. Fulde, and H. Stoll,
Phys. Rev. B {\bf 62}, 5482 (2000).

\bibitem{Harl2008}
J.~Harl and G.~Kresse,
Phys. Rev. B {\bf 77}, 045136 (2008).

\bibitem{Peterson1966}
O. G. Peterson, D. N. Batchelder, and R. O. Simmons,
Phys. Rev. {\bf 150}, 703 (1966).

\bibitem{Schwalbe1977}
L. A. Schwalbe, R. K. Crawford, H. H. Chen, and R. A. Aziz, 
J. Chem. Phys. {\bf 66}, 4493 (1977).

\bibitem{Wittlinger1997}
J. Wittlinger, R. Fischer, S. Werner, J. Schneider, and H. Schulz,
Acta Cryst. B {\bf 53}, 745 (1997).

\bibitem{Ishikawa2014}
T. Ishikawa, M. Asano, M. Obata, N. Suzuki, T. Oda, I. Hamada, and K. Shimizu, 
private communication.

\bibitem{Spanu2009}
L.~Spanu, S.~Sorella, and G.~Galli,
Phys. Rev. Lett. {\bf 103}, 196401 (2009).

\bibitem{Lebegue2010}
S. Leb\`{e}gue, J. Harl, T. Gould, J. G. \'{A}ngy\'{a}n, G. Kresse, and J. F. Dobson,
Phys. Rev. Lett. {\bf 105}, 196401 (2010).

\bibitem{Bjoerkman2012}
T.~Bj{\"o}rkman, A.~Gulans, A.~V.~Krasheninnikov, and R.~M.~Nieminen,
Phys. Rev. Lett. {\bf 108}, 235502 (2012).

\bibitem{Baskin1955}
Y. Baskin and L. Meyer,
Phys. Rev. {\bf 100}, 544 (1955).

\bibitem{Zacharia2004}
R. Zacharia, H. Ulbricht, and T. Hertel,
Phys. Rev. B {\bf 69}, 155406 (2004).

\bibitem{Hamada2010}
I.~Hamada and M.~Otani,
Phys. Rev. B {\bf 82}, 153412 (2010).

\bibitem{Mapasha2012}
R.~E.~Mapasha, A.~M.~Ukpong, and N.~Chetty,
Phys. Rev. B {\bf 85}, 205402 (2012).

\bibitem{Graziano2012}
G.~Graziano, J.~Klime\v{s}, F.~Fernandez-Alonso, and A.~Michaelides,
J. Phys.: Condens. Matter {\bf 24}, 424216 (2012).

\bibitem{JuNoLo}
rVV10 was also implemented in the JuNoLo code \cite{Lazic2010} using the Rom\'{a}n-P\'{e}rez-Soler (RPS) algorithm.
The implementation of the RPS algorithm is described by Callsen {\it et al}.\cite{Callsen2012}

\bibitem{Lazic2010}
P.~Lazi\'{c}, N.~Atodiresei, M.~Alaei, V.~Caciuc, S.~Bl\"{u}gel, and R.~Brako,
Comput. Phys. Commun. {\bf 181}, 371 (2010).

\bibitem{Callsen2012}
M.~Callsen, N.~Atodiresei, V.~Caciuc, and S.~Bl\"{u}gel,
Phys. Rev. B {\bf 86}, 085439 (2012).

\bibitem{Callsen2014}
M.~Callsen and I.~Hamada,
private communication.

\bibitem{Swanson1955}
H. E. Swanson, N. T. Gilfrich, and G. M. Ugrinic , 
National Bureau of Standerds Circular 539, Vol. 5 
(U. S. Government Printing Office, Washington, D. C., 1955) p.~54.

\bibitem{Cherin1967}
P. Cherin and P. Unger, 
Inorg. Chem. {\bf 6}, 1589 (1967). 

\bibitem{DalCorso1994}
A. D. Corso and R. Resta, 
Phys. Rev. B {\bf 50}, 4327 (1994). 

\bibitem{Kresse1994}
G.~Kresse, J. Furthm{\" u}ller, and J.~Hafner,
Phys. Rev. B {\bf 50}, 13181 (1994).

\bibitem{Bucko2010}
T.~Bu\v{c}ko, J.~Hafner, S.~Leb\`{e}gue, and J.~G.~\'{A}ngy\'{a}n,
J. Phys. Chem. A {\bf 114}, 11814 (2010).

\bibitem{Sode2013}
O. Sode, M. Ke\c{c}, K. Yagi, and S. Hirata,
J. Chem. Phys. {\bf 138}, 074501 (2013).

\bibitem{Simon1980}
A. Simon and K. Peters,
Acta Cryst. B {\bf 36}, 2750 (1980).

\bibitem{Otero-de-la-Roza2012}
A.~Otero-de-la-Roza and E.~R.~Johnson,
J. Chem. Phys. {\bf 137}, 054103 (2012).

\bibitem{Bonev2003}
S. A. Bonev, F. Gygi, T. Ogitsu, and G. Galli,
Phys. Rev. Lett. {\bf 91}, 065501 (2003).

\bibitem{Hamada2013}
I.~Hamada,
private communication.

\bibitem{Corsetti2013}
F.~Corsetti, E.~Artacho, J.~M.~Soler, S.~S.~Alexandre, and M.-V.~Fernandez-Serra,
J. Chem. Phys. {\bf 139}, 194502 (2013).

\end{thebibliography}
\end{document}